\documentstyle[aps,prl,twocolumn,graphicx,floats]{revtex}

\newcommand{\be}{\begin{eqnarray}}
\newcommand{\ee}{\end{eqnarray}}

\begin{document}
\draft
\twocolumn[\hsize\textwidth\columnwidth\hsize\csname @twocolumnfalse\endcsname
\title{A Scaling Behavior of Spectral Weight Changes in Perovskite Manganites La$%
_{0.7-{\it {y}}}$Pr$_{{\it y}}$Ca$_{0.3}$MnO$_3$}
\author{K. H. Kim$^1$, J. H. Jung$^1$, D. J. Eom$^1$, T. W. Noh$^{1,*}$, Jaejun Yu$^2
$, and E. J. Choi$^3$}
\address{$^1$Department of Physics, Seoul National University, Seoul 151-742, Korea}
\address{$^2$Department of Physics, Sogang University, Seoul 121-742, Korea}
\address{$^3$Department of Physics, Seoul City University, Seoul 130-743, Korea}
\date{\today }
\maketitle

\begin{abstract}
~Optical conductivity spectra of La$_{0.7-{\it {y}}}$Pr$_{{\it y}}$Ca$_{0.3}$%
MnO$_3$ were systematically investigated. For metallic samples, the spectral
weight below 0.5 eV, whose magnitude can be represented by the effective
carrier number $N_{{\rm {eff}}}$(0.5 eV), increases as temperature becomes
lower. Regardless of the Pr doping, all the measured values of $N_{{\rm {eff}%
}}$(0.5 eV)/$T_C$ fall into one scaling curve. This scaling behavior could
be explained by the theoretical model by R\"{o}der {\it et al}. $\left( 
\text{Phys. Rev. Lett. {\bf 76}, 1356 (1996)}\right) $, which includes spin
double exchange and Jahn-Teller lattice coupling to holes. With the Pr
doping, far-infrared conductivities were found to be suppressed, probably
due to the Anderson localization.
\end{abstract}

\pacs{PACS number; 72.15.Gd, 75.50.Cc, 75.30.Kz, 78.20.Ci}

\vskip1pc] 

Physics of doped manganites, {\it R}$_{1-x}${\it A}$_x$MnO$_3$ ({\it R}=rare
earth and {\it A}=alkaline earth ions) with 0.2$\leq ${\it x}$\leq $0.5, has
been investigated extensively since the recent discovery of colossal
magnetoresistance (CMR) phenomena in these compounds. Even if the double
exchange (DE) interaction can qualitatively explain most of their physical
properties\cite{DE}, a lot of recent works have revealed that some
additional degrees of freedom should be included to describe them more
accurately\cite
{Okimoto1,Okimoto2,Nagaosa,Shiba,Millis1,Millis2,Kaplan,kim96,kim98,Zhang}.
Optical techniques have been useful to address such additional degrees of
freedom\cite{Okimoto1,Okimoto2,Kaplan,kim96,kim98}.

However, there still remain some controversies on how to explain temperature(%
{\it T})-dependent spectral weight (SW) changes in optical spectra of the
doped manganites. Especially, there have been numerous interpretations on
the mid-infrared (IR) absorption peak below 0.5 eV, which appears below the
Curie temperature {\it T}$_{{\rm {C}}}$ and becomes stronger at lower
temperatures. Okimoto {\it et al.} claimed that the SW changes should be
understood in the spin-split band picture, and they assigned the mid-IR
feature to an intraband excitation within an {\it e}$_g$ band which was
merged below {\it T}$_{{\rm {C}}}$ from two spin-split bands\cite
{Okimoto1,Okimoto2}. Other workers suggested that an orbital degree of
freedom should be included to explain the SW changes\cite{Nagaosa,Shiba},
and de Brito and Shiba attributed the mid-IR absorption to an interband
transition between the {\it e}$_g$ orbital states\cite{Shiba}. Millis {\it %
et al.} showed that the dynamic Jahn-Teller (JT) interaction could play an
important role in the SW changes\cite{Millis1,Millis2}, and Kaplan {\it et
al.} assigned the mid-IR feature to a JT type small polaron absorption\cite
{Kaplan}. Recently, through detailed studies on optical properties of La$%
_{0.7}$Ca$_{0.3}$MnO$_3$ (LCMO), we proposed that the evolution of the low
frequency feature below 0.5 eV come from a crossover from small to large
polaron states\cite{kim98}. In other words, the mid-IR feature should be
attributed to an incoherent absorption of a large polaron state, whose
existence was predicted by R\"{o}der {\it et al.}\cite{Zhang}.

To get further insights on this interesting issue, we investigated optical
properties of La$_{0.7-y}$Pr$_y$Ca$_{0.3}$MnO$_3$ (LPCMO). Hwang {\it et al.}
showed that dc transport properties of LPCMO were affected by a carrier
hopping parameter which could be varied systematically by the Pr doping\cite
{hwang95}. In this paper, we report on optical conductivity spectra of the
LPCMO samples with {\it y}=0.13, 0.4, 0.5, and 0.7. We found that detailed
conductivity spectra vary significantly with the Pr doping. However, the 
{\it T}-dependence of the SW below 0.5 eV remains similar. Actually, when
all of the {\it T}-dependent SW are scaled by corresponding {\it T}$_{{\rm {C%
}}}$ values, they fall into one scaling curve. This scaling behavior will be
explained using the theoretical results by R\"{o}der {\it et al.}\cite{Zhang}%
.

Polycrystalline LPCMO samples were prepared by a standard solid-state
reaction method.~The samples were oxygen-annealed at 1100 $^{\circ }$C for
72 hours\cite{kim97}. X-ray diffraction and electron-probe microanalysis
measurements confirmed that the samples were single-phase and
stoichiometric. Both resistivity and magnetization curves showed hysteretic
behaviors, confirming that corresponding phase transitions are of
first-order nature\cite{hwang95}. For the samples with {\it y}=0.13, 0.4,
and 0.5, the metal-insulator (M-I) transition temperatures, defined by
temperatures of resistivity maxima in warming (cooling) runs, were 241
(239), 156 (152), and 123 K (118 K), respectively. The {\it y}=0.7 sample
did not show any M-I transition. Values of $T_C$ were estimated from
field-cooled dc magnetization curves, which were measured at 100 Oe using a
commercial SQUID magnetometer. Due to the hysteretic behaviors, onset
temperatures of the dc magnetization for warming and cooling runs were
measured and averaged. Values of $T_C$ for the $y$=0.13, 0.4, and 0.5
samples were found to be 238$\pm $2, 155$\pm $3, and 120$\pm $3 K,
respectively.

Near normal incidence reflectivity spectra, $R(\omega )$, were measured
between 0.01 and 30 eV. A liquid He-cooled cryostat was used to measure {\it %
T-}dependence of $R(\omega )$ in a frequency region between 0.01 and 2.5 eV.
Optical conductivity spectra, $\sigma $($\omega $), were obtained using the
Kramers-Kronig transformation with appropriate extrapolations\cite{kim98}.

Figure \ref{Fig:PLcondm} shows $\sigma $($\omega $) of the LPCMO samples
below 2.0 eV. Note that $\sigma $($\omega $) of all the samples are very
similar at 290 K: each of $\sigma $($\omega $) has a broad absorption
feature centered around 1 eV and vanishes as $\omega $ approaches to zero.
These gap-like features are consistent with the fact that these samples are
in insulating states at room temperature. For Pr$_{{\rm {0.7}}}$Ca$_{{\rm {%
0.3}}}$MnO$_{{\rm {3}}}$, $\sigma $($\omega $) are nearly  {\it T}%
-independent, since it remains in an insulating state down to 15 K. For the
samples with {\it y}=0.13, 0.4 and 0.5, $\sigma $($\omega $) show quite
strong $T$-dependence. As {\it T} decreases below {\it T}$_{{C}}$, the
spectral features approximately above 0.5 eV decrease and those below 0.5 eV
increase. Interestingly, for the {\it y}=0.4 sample, such SW changes at all
temperatures occur around one balancing point, fixed near 0.5 eV\cite
{isosbetic}.

Note that some of $\sigma $($\omega $) show a double peak structure. For
example, the {\it y}=0.4 sample displays two peaks centered around 0.25 and
1.5 eV at 120 K. The double peak feature is quite similar to that of two
mid-gap states observed in the doping dependent $\sigma $($\omega $) of La$%
_{1-x}$Ca$_x$MnO$_3$\cite{Jung2}. Based on this similarity, we attribute the
absorption band near 1.5 eV to an intra-atomic transition between JT split $%
e_g$ states\cite{Jung2,Jung1}, and the mid-IR absorption peak below 0.5 eV
to an interatomic charge transfer transition between Mn$^{3+}$({\it e}$_g^1$%
) and Mn$^{4+}$ ({\it e}$_g$) with the same {\it t}$_{2g}$ background spins%
\cite{Millis2,Kaplan,Jung2}. In our recent paper on LCMO, we proposed that
the mid-IR band below 0.5 eV come from incoherent absorption of a large
polaron\cite{kim98}.

In Fig.~\ref{Fig:PLcondm}, as {\it y} increases, the mid-IR absorption
features decrease systematically. To get further understanding on the mid-IR
absorption band, an effective carrier number, $N_{eff}(\omega _{{\rm {c}}})$%
, of carriers below a cut off frequency, $\omega _{{\rm {c}}}$, was
evaluated using the following relation: 
\begin{equation}
N_{eff}(\omega _c)=\frac{2m}{\pi e^2N}\int_0^{\omega _c}\sigma (\omega
)d\omega {\rm {,}}
\end{equation}
where {\it m} is an electron mass of carriers and {\it N} is a number of Mn
ions per unit volume. For actual evaluation, the value of $\hbar \omega _{%
\text{c}}$ was chosen as 0.5 eV\cite{cutoff}. And, in evaluating {\it N, T}%
-dependent volume change was considered properly from reported structural
data\cite{radaelli97}.

Figure~\ref{Fig:PLneff} shows $N_{eff}$($\omega _{\text{c}}$) vs {\it T}
curves for {\it y}=0.13 (solid circles), 0.4 (solid squares), and 0.5
(crosses). For comparison, values of $N_{eff}$($\omega _{\text{c}}$) for the
LCMO sample (open diamonds) are also included\cite{kim98}. It is found that
the $N_{eff}$($\omega _{\text{c}}$) vs {\it T} curve for LCMO is quite close
to that of the {\it y}=0.13 sample. [The $T_{\text{C}}$ value of LCMO is 245$%
\pm $2 K, which is close to that of the {\it y}=0.13 sample.] Note that $%
N_{eff}$($\omega _{\text{c}}$) for each sample increases abruptly near $T_{%
\text{C}}$. As the Pr doping increases, $N_{eff}$($\omega _{\text{c}}$)
decreases systematically at overall temperatures.

In Fig.~\ref{Fig:PLneff}, it is interesting to note that the shape of $%
N_{eff}$($\omega _{\text{c}}$) for each sample is quite similar. And, the
increasing rate of $N_{eff}$($\omega _{\text{c}}$) is nearly as same as that
of $T_{\text{C}}$: for example, when $T_{\text{C}}$ is reduced by about a
factor of 2, $N_{eff}$($\omega _{\text{c}}$) is also reduced by a similar
factor. So, we plotted {\it N}$_{eff}$($\omega _c$)/$T_{{\rm {C}}}$ vs {\it T%
}/$T_{{\rm {C}}}$ curves, shown in Fig.~\ref{Fig:PLscaling}. Interestingly
enough, even if detailed SW transfer behaviors in Fig. 1 are quite
different, all of experimental values of {\it N}$_{eff}$($\omega _c$)/$T_{%
{\rm {C}}}$ fall into nearly in one curve. The result demonstrates that the 
{\it N}$_{eff}$($\omega _c$)/$T_{{\rm {C}}}$ vs {\it T}/$T_{{\rm {C}}}$
curves in these LPCMO compounds can be described by one scaling function.
Moreover, the scaling function is very close to $\gamma _{{\rm {DE}}}$({\it T%
}), which is the {\it T}-dependent DE bandwidth predicted by Kubo and Ohata%
\cite{Kubo}. The solid line in Fig. \ref{Fig:PLscaling} displays the
behavior of $\gamma _{{\rm {DE}}}$({\it T/}$T_{{\rm {C}}}$).

To explain this scaling behavior in the {\it N}$_{eff}$($\omega _c$)/$T_{%
{\rm {C}}}$ vs {\it T}/$T_{{\rm {C}}}$ curves, we adopt theoretical results
by R{\"{o}}der {\it et al.}\cite{Zhang}, who investigated a Hamiltonian
including the DE interaction and the JT electron-phonon coupling terms. [A
similar Hamiltonian was also investigated by Millis {\it et al.}\cite
{Millis1,Millis2}.] After the Lang-Firsov transformation for the
electron-phonon problem, the Hamiltonian could be reduced to an effective
Hamiltonian over a phonon vacuum: 
\begin{equation}
\widetilde{H}=-t\xi (\eta )\sum_{<ij>\sigma }\cos \frac{\theta _{ij}}2\left(
c_{i\sigma }^{\dagger }c_{j\sigma }+H.c.\right) +\sum_iD_i\;,
\label{Eq:JTScaling}
\end{equation}
where {\it t} is the bare hopping matrix element. The polaronic band
narrowing is given by $\xi $ ($\eta $)=exp($-\epsilon _p\eta ^2$/$\hbar
\omega _0$) with an effective electron-phonon coupling parameter $\epsilon
_p $ and a coupled phonon frequency $\omega _0$. Here, $\eta $ measures the
degree of the polaron effect, which is determined self-consistently and
dependent on $\epsilon _p$, $\omega _0$, and the hole carrier concentration, 
{\it x}. Note that $\xi $ ($\eta $) is quite sensitive to a variation of 
{\it t} since $\epsilon _p$ is inversely proportional to {\it t}. {\it D}$_i$
refers to diagonalized terms over the phonon vacuum. [For zero
electron-phonon coupling, $\xi $($\eta $) simply becomes 1.] Within the
Hamiltonian given in Eq. (2), R{\"{o}}der {\it et al.}\cite{Zhang} derived 
\begin{equation}
T_C(x)=\frac 9{50}[-e_B(x)]\xi (\eta ),  \label{Eq:JT-Tc}
\end{equation}
where $e_B$($x$) is a band energy, which is proportional to {\it t }and
depends on $x$. As $\eta $ approaches zero, $e_B$($x$) is roughly
proportional to {\it x}(1$-${\it x}).

If we assume that the mid-IR absorption below 0.5 eV should be proportional
to the hopping term in Eq. (2),

\begin{equation}
{\it N}_{eff}(\omega _c;{\it T})\approx {\it t}\xi (\eta )\langle \cos
(\theta /2)\rangle \approx {\it t}\xi (\eta )\gamma _{{\rm {DE}}}({\it T}).
\label{Eq:Neff}
\end{equation}
In the latter approximation, the spin-dependent cosine term was averaged
thermodynamically and replaced with the mean field result, i.e. $\gamma _{%
{\rm {DE}}}$({\it T}). Note that, in Eqs. (3) and (4), both ${\it N}%
_{eff}(\omega _c;{\it T})$ and $T_C$ include the same factor $\xi $($\eta $%
), which represents the polaronic band narrowing effect\cite{Mahan,JDLee}.
For all the LPCMO samples, the hole concentration can be considered to be
fixed, i.e. 0.3 per Mn atom, so {\it N}$_{eff}$($\omega $$_c$;{\it T})/$T_C$
should be proportional to $\gamma _{{\rm {DE}}}$({\it T}). Indeed, the inset
of Fig. \ref{Fig:PLscaling} shows such a linear relationship.

As far as we know, the theory by R{\"{o}}der {\it et al.}\cite{Zhang} is the
only model that can explain the scaling behavior. In this model, both $T_C$
and the interatomic transfer matrix between Mn$^{3+}$({\it e}$_g^1$) and Mn$%
^{4+}$ ({\it e}$_g$) will be significantly affected at the same time by
changes in the polaronic band narrowing factor. The variation of {\it t} was
estimated to be less than 2 \% in the LPCMO compounds under the
tight-binding approximation\cite{radaelli97}. Note that the small change in 
{\it t} is difficult to explain the large variations in ${\it N}%
_{eff}(\omega _c;{\it T})$ and $T_C$ without the self-amplifying coupling of
both DE and JT effects\cite{Zhang}. Moreover, R{\"{o}}der {\it et al.}\cite
{Zhang} predicted that a large polaron state should exist at a low
temperature and that such an electron-phonon interaction should be also
coupled with the spin degree of freedom. If such a large polaron state
exists, as we demonstrated in our earlier paper\cite{kim98}, it should be in
a strong electron-phonon coupling limit, which has not been realized in
other real system before.

In Fig. \ref{Fig:PLscaling}, there are small discrepancies between
experimental data and $\gamma _{{\rm {DE}}}$({\it T}). For the samples of 
{\it y=}0.4 and 0.5, the experimental data are slightly deviated from $%
\gamma _{{\rm {DE}}}$({\it T}) near {\it T}$_C$: the increase of {\it N}$%
_{eff}$($\omega _c$) near {\it T}$_C$ is rather gradual. [This behavior
might be related to the dc resistivity behavior, showing a broader M-I
transition for a sample with a larger Pr doping\cite{hwang95}.] Note that
the {\it y=}0.7 sample, i.e. Pr$_{{\rm {0.7}}}$Ca$_{{\rm {0.3}}}$MnO$_{{\rm {%
3}}}$, is a canted antiferromagnetic insulator\cite{kawano}. Therefore, the
observed discrepancies might be related to enhanced spin fluctuations near $%
T_C$ due to the increased antiferromagnetic spin interaction relative to the
ferromagnetic one.

As shown in Fig.~1, below {\it T}$_C$, a broad SW between 0.5 and 1.5 eV
starts to decrease and a Drude peak and a broad mid-IR peak below 0.5 eV
starts to appear. In the large polaron picture, these SW changes can be
interpreted such that there is a collapse of the JT small polaron and a
gradual crossover to the large polaron state as the sample enters the
ferromagnetic metallic state. However, quite an unusual behavior was
observed in the far-IR spectra.

Figure~4 shows the far-IR $\sigma $($\omega $) of LPCMO. Estimated dc
conductivity values from the Hagen-Rubens relation\cite{kim97}, which were
denoted as symbols, agree reasonably well with the extrapolated values of $%
\sigma $($\omega $) as $\omega \rightarrow 0$. For a metallic sample, as $%
\omega $ decreases, $\sigma $($\omega $) increases initially but shows a
downturn at a very low frequency. Such a trend becomes more evident, as {\it %
y} increases, i.e. as the carrier hopping parameter is reduced. [For a pure
LCMO sample, we could not observe such an effect up to a measured frequency
limit of 8 meV\cite{kim98}.] Therefore, the Drude peak seems to disappear
systematically with increasing the Pr doping, even if the sample shows a
metallic transport behavior below {\it T}$_{{\rm {C}}}$. This observation
suggests that the LPCMO samples with large Pr doping, i.e. {\it y=}0.4 and
0.5, are far from normal Fermi liquids at low {\it T}. Similar behaviors of $%
\sigma $($\omega $) in the far-IR region were also observed in a layered
manganites\cite{Ishikawa} and La$_{1-y}$TiO$_{3-\delta }$\cite{timusk}.

We postulate that the unusual far-IR response of $\sigma $($\omega $) be
caused by the Anderson localization effects. Many theoretical works
predicted that disorder effects could be important in the CMR manganites\cite
{Sheng97}. For the LPCMO samples, there are several possible causes which
bring out randomness in the effective hopping matrix: (1) Mn-O-Mn bond angle
and/or length disorder, (2) inhomogeneous JT lattice distortions, (3)
spin-exchange interaction between the Pr and the Mn ions, and (4)
competition between the ferromagnetic DE and the antiferromagnetic
superexchange interaction among the Mn sites\cite{coey}. Such randomness
sources can produce a mobility edge which is located close to the Fermi
energy, resulting in the suppression of the Drude peak. And, for a sample
with a smaller value of the carrier hopping term, i.e. a higher Pr doping,
its carrier transport will be affected more easily by the mobility edge, so
the Anderson localization effects should be more significant in its far-IR
conductivity spectrum.

In summary, we investigated the SW changes, especially in the mid-IR region,
of the La$_{0.7-y}$Pr$_y$Ca$_{0.3}$MnO$_3$ samples. The Pr doping results in
suppression of the far-IR conductivity, probably due to the Anderson
localization. It was also found that the experimental data of {\it N}$_{eff}$%
($\omega $$_c$)/$T_C$ could be scaled into a curve, which is linearly
proportional to $\gamma _{{\rm {DE}}}$({\it T}), regardless of the Pr doping
value. This interesting scaling behavior was explained in terms of the
theoretical model by R\"{o}der {\it et al}., which takes into account of a
JT coupling in the DE model. This work clearly shows the importance of the
lattice effects in the optical properties of the CMR manganites.

One of us (TWN) appreciate the hospitality of the LG-CIT during his visit
when this manuscript was written.
We acknowledge the financial support by 
Ministry of Education through the grant No.~BSRI-97-2416 and by the Korea
Science and Engineering Foundation through the grant No.~96-0702-02-01 and
through RCDAMP of Pusan National University. Reflectivity measurements in
the vacuum ultraviolet region were performed at Pohang Light Source,
supported by POSCO and Ministry of Science and Technology.

$^{*}$Also at LG Corporate Institute of Technology (LG-CIT), Seoul, Korea.
Electronic mail address: twnoh@phya.snu.ac.kr.

\newpage

\begin{figure}
\caption{Optical conductivity spectra of La$_{0.7-y}$Pr$_y$Ca$_{0.3}$MnO$_3$
below 2.0 eV. }
\label{Fig:PLcondm}
\end{figure}

\begin{figure}
\caption{$T$-dependence of $N_{eff}$ ($\omega _{\text{c}}$ = 0.5 eV) 
for La$_{0.7-y}$Pr$_y$Ca$_{0.3}$MnO$_3$. }
\label{Fig:PLneff}
\end{figure}

\begin{figure}
\caption{$N_{eff}$ ($\omega _c$ = 0.5 eV)/$T_C$ vs $T$/$T_{{\rm {C}}}$
curves of La$_{0.7-y}$Pr$_y$Ca$_{0.3}$MnO$_3$. The solid line refers to the
behavior of $\gamma _{{\rm {DE}}}$({\it T})$\equiv \langle \cos (\theta
/2)\rangle $. The inset shows a linear scaling behavior between $N_{eff}$ ($%
\omega _c$ = 0.5 eV)/$T_C$ and $\gamma _{{\rm {DE}}}$({\it T}).}
\label{Fig:PLscaling}
\end{figure}

\begin{figure}
\caption{Optical conductivity spectra of 
La$_{0.7-y}$Pr$_y$Ca$_{0.3}$MnO$_3$
in the far-infrared region below 0.15 eV. Symbols denote the dc conductivity
values estimated from the Hagen-Rubens relation.}
\label{Fig:PLcondf}
\end{figure}

\newpage
\clearpage
\vspace*{0cm}
\begin{center}
\hspace*{1cm}
\includegraphics[height=16cm,width=15cm,angle=180]{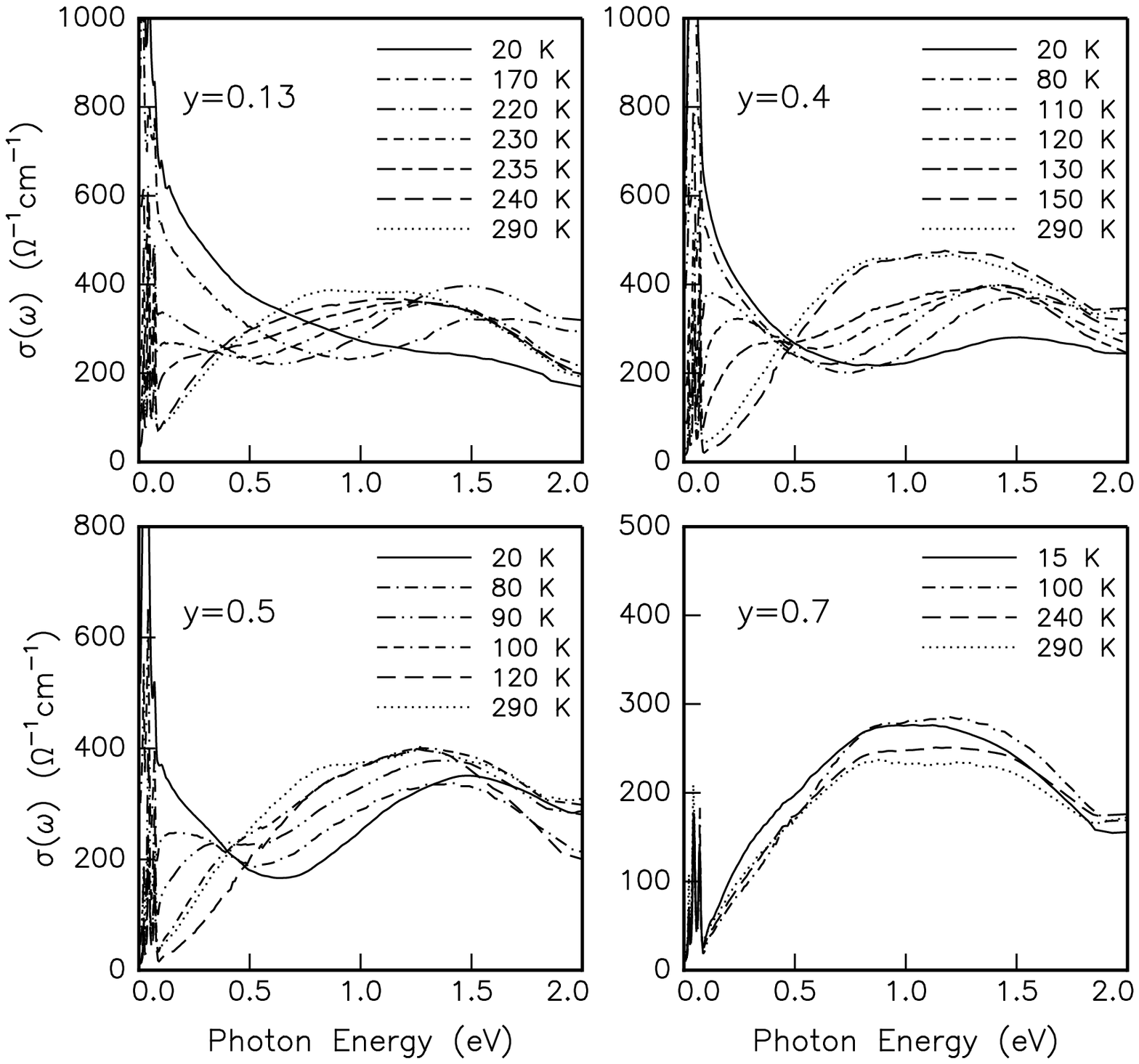}
\end{center}
\vspace*{3.0cm}
\large
\hspace{1cm} FIG. 1. K. H. Kim {\it et al.} 

\newpage
\clearpage
\vspace*{0cm}
\begin{center}
\hspace*{1cm}
\includegraphics[height=15cm,width=16cm,angle=180]{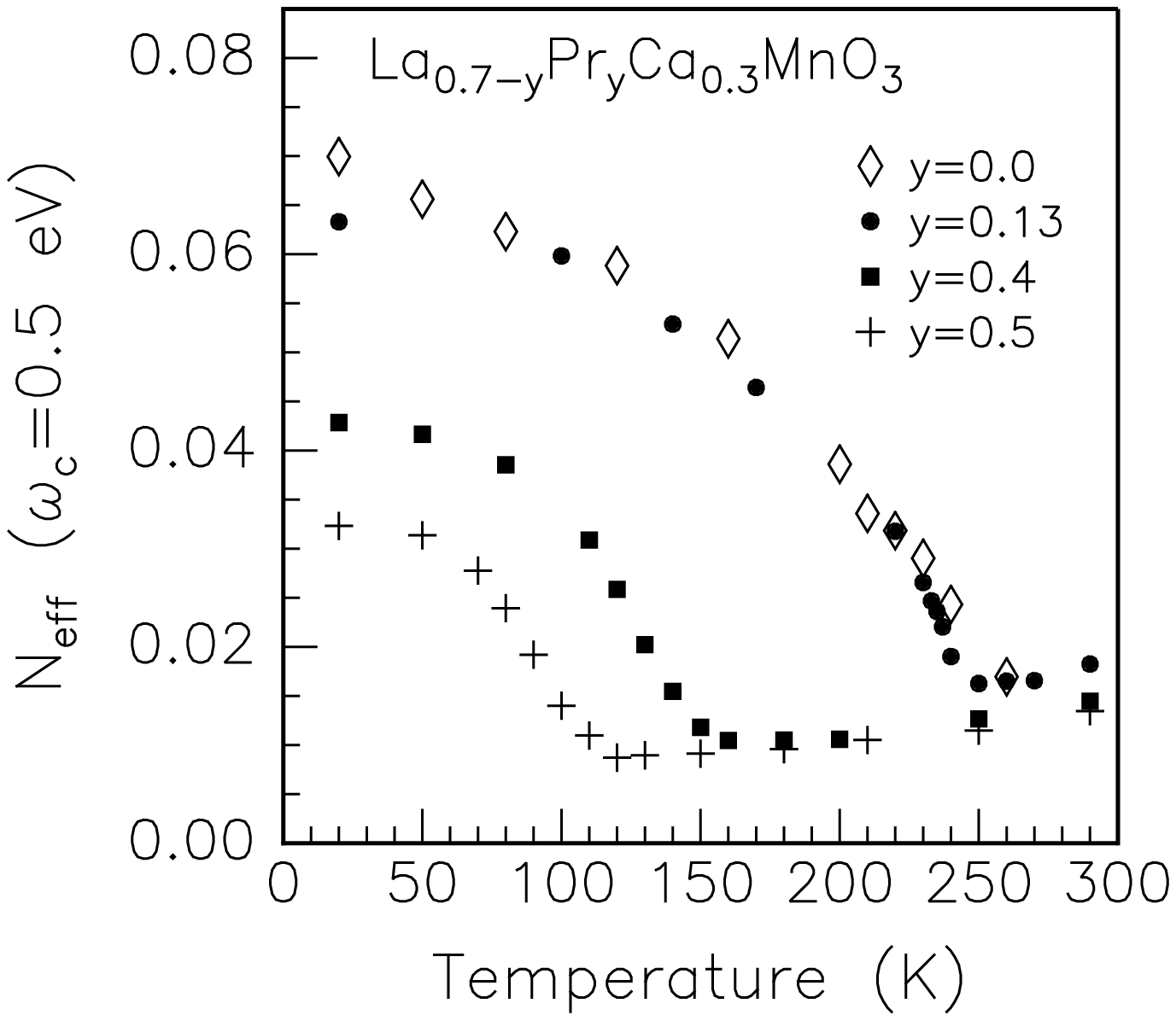}
\end{center}
\vspace*{3.0cm}
\large
\hspace{1cm} FIG. 2.  K. H. Kim {\it et al.}

\newpage
\clearpage
\vspace*{0cm}
\begin{center}
\hspace*{1cm}
\includegraphics[height=15cm,width=16cm,angle=180]{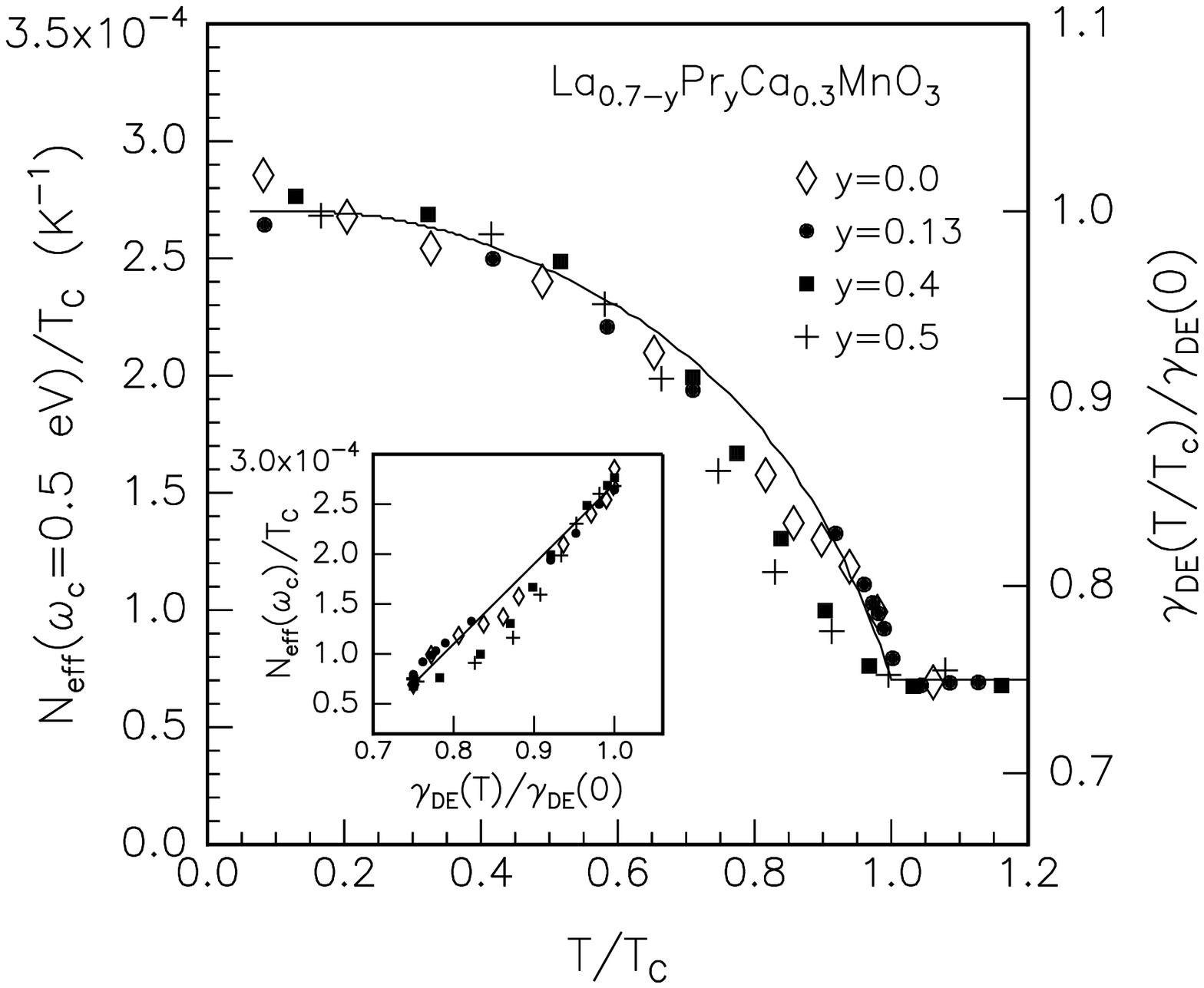}
\end{center}
\vspace*{3.0cm}
\large
\hspace{1cm} FIG. 3.  K. H. Kim {\it et al.}

\newpage
\clearpage
\vspace*{0cm}
\begin{center}
\hspace*{1cm}
\includegraphics[height=16cm,width=15cm,angle=180]{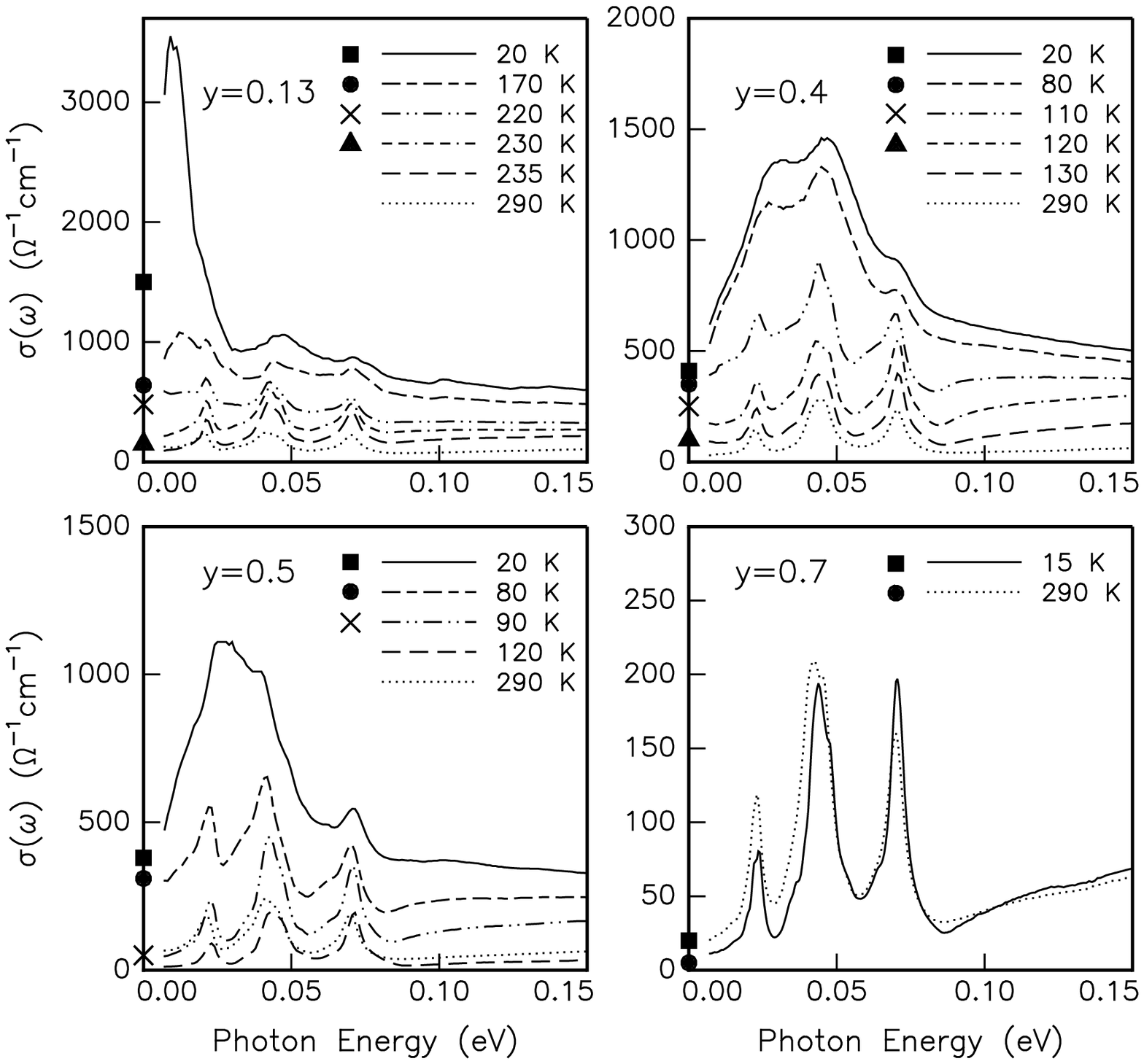}
\end{center}
\vspace*{3.0cm}
\large
\hspace{1cm} FIG. 4.  K. H. Kim {\it et al.}


\begin{references}
\bibitem{cutoff}  Even if $\omega _c$ is chosen with a slightly different
value, e.g. 0.6 eV, the shape of the SW changes and their scaling behaviors
are nearly the same with those shown in Figs. 2 and 3.
\end{references}
\end{document}